\documentclass[pdflatex,sn-mathphys-num]{sn-jnl}

\usepackage{graphicx}%
\usepackage{multirow}%
\usepackage{amsmath,amssymb,amsfonts}%
\usepackage{amsthm}%
\usepackage{mathrsfs}%
\usepackage[title]{appendix}%
\usepackage{xcolor}%
\usepackage{textcomp}%
\usepackage{manyfoot}%
\usepackage{booktabs}%
\usepackage{algorithm}%
\usepackage{algorithmicx}%
\usepackage{algpseudocode}%
\usepackage{listings}%
\usepackage{bm}
\usepackage{color}

\definecolor{codegreen}{rgb}{0,0.6,0}
\definecolor{codegray}{rgb}{0.5,0.5,0.5}
\definecolor{codepurple}{rgb}{0.58,0,0.82}
\definecolor{backcolour}{rgb}{0.95,0.95,0.92}

\lstdefinestyle{mystyle}{
  backgroundcolor=\color{backcolour}, commentstyle=\color{codegreen},
  keywordstyle=\color{magenta},
  numberstyle=\tiny\color{codegray},
  stringstyle=\color{codepurple},
  basicstyle=\ttfamily\scriptsize,
  language=Python,
  morekeywords={True, False},
  frame=none,
  breakatwhitespace=false,         
  breaklines=true,
  captionpos=b,                    
  keepspaces=true,                 
  numbers=none,                    
  numbersep=5pt,                  
  showspaces=false,                
  showstringspaces=false,
  showtabs=false,                  
  tabsize=4,
  float
}

\lstdefinestyle{lmpstyle}{
  backgroundcolor=\color{backcolour}, commentstyle=\color{codegreen},
  basicstyle=\ttfamily\scriptsize,
  frame=none,
  breakatwhitespace=false,         
  breaklines=true,                 
  captionpos=b,                    
  keepspaces=true,                 
  numbers=none,                    
  numbersep=5pt,                  
  showspaces=false,                
  showstringspaces=false,
  showtabs=false,                  
}

\theoremstyle{thmstyletwo}%

\theoremstyle{thmstylethree}%

\newcommand{\ctd}{{\ttfamily chemtrain-deploy}}

\newcommand{\sref}[2]{\hyperref[#1]{\ref*{#1}#2}}

\begin{document}

\title[Article Title]{\texttt{chemtrain-deploy:} A parallel and scalable framework for machine learning potentials in million-atom MD simulations.}


\author[1]{\fnm{Paul} \sur{Fuchs}}

\author[1]{\fnm{Weilong} \sur{Chen}}

\author[3]{\fnm{Stephan} \sur{Thaler}}

\author*[1,2]{\fnm{Julija} \sur{Zavadlav}}\email{julija.zavadlav@tum.de}

\affil[1]{Professorship of Multiscale Modeling of Fluid Materials, Department of Engineering Physics and Computation, TUM School of Engineering
and Design, Technical University of Munich, Germany}

\affil[2]{Atomistic Modeling Center (AMC), Munich Data Science Institute (MDSI), Technical University of Munich, Germany}

\affil[3]{Valence Labs, Montreal, QC, Canada}

\abstract{Machine learning potentials (MLPs) have advanced rapidly and show great promise to transform molecular dynamics (MD) simulations. However, most existing software tools are tied to specific MLP architectures, lack integration with standard MD packages, or are not parallelizable across GPUs. To address these challenges, we present \texttt{chemtrain-deploy}, a framework that enables model-agnostic deployment of MLPs in LAMMPS. \texttt{chemtrain-deploy} supports any JAX-defined semi-local potential, allowing users to exploit the functionality of LAMMPS and perform large-scale MLP-based MD simulations on multiple GPUs. It achieves state-of-the-art efficiency and scales to systems containing millions of atoms. We validate its performance and scalability using graph neural network architectures, including MACE, Allegro, and PaiNN, applied to a variety of systems, such as liquid–vapor interfaces, crystalline materials, and solvated peptides. Our results highlight the practical utility of \texttt{chemtrain-deploy} for real-world, high-performance simulations and provide guidance for MLP architecture selection and future design.}

\maketitle

\section{Introduction}\label{sec1}
In recent years, machine learning potentials (MLPs) have advanced rapidly and found widespread applications in fields such as computational chemistry and materials science~\cite{unkeMachineLearningForce2021a, noeMachineLearningMolecular2020, behler2016perspective, merchantScalingDeepLearning2023}. By training the models on high-accuracy reference datasets, typically consisting of energies and forces, MLPs offer a promising compromise between accuracy and computational efficiency. They approach the accuracy of ab initio methods~\cite{iftimieInitioMolecularDynamics2005} while maintaining computational speeds closer to classical force fields~\cite{cornellSecondGenerationForce1995, nikitinNewAMBERcompatibleForce2014, marrinkCoarseGrainedModel2004}. This performance is achieved by capturing high-order, many-body interactions through either predefined~\cite{behlerAtomcenteredSymmetryFunctions2011, smithANI2017} or learned representations of local atomic environments~\cite{gilmerMessagePassingGNN2017, drautzAtomicClusterExpansion2019}, enabling near-linear scaling with system size~\cite{zhangDeepPotentialMolecular2018a}, and making them particularly attractive for large-scale simulations.

While there have been significant advancements~\cite{rhodesOrbv3AtomisticSimulation2025, zhangGraphNeuralNetwork2025, kovacsMACEOFFShortRangeTransferable2025}, several challenges still hinder the widespread adoption of MLPs in real-world applications. From a model architecture perspective, graph neural networks (GNNs) have emerged as a powerful approach due to their natural ability to encode atomic topologies and interactions~\cite{batatiaDesignSpaceE3equivariant2025, musaelianAllegro2023, batznerNequIP2022, schuttSchNet2018, schuttEquivariantMessagePassing2021, batatiaMACEHigherOrder2023, zhangGraphNeuralNetwork2025}. Numerous GNN architectures leveraging geometric priors have been proposed, with many accompanied by specialized software packages such as SchNetPack~\cite{schuttSchNetPackDeepLearning2019}, TorchANI~\cite{gaoTorchANIFreeOpen2020}, MACE~\cite{batatiaMACEHigherOrder2023}, TorchMD~\cite{doerrTorchMDDeepLearning2021} and DistMLIP~\cite{hanDistMLIPDistributedInference2025}. However, these packages are often tightly coupled to their respective models and typically lack the modularity or plugin interfaces needed for seamless integration with widely used molecular dynamics (MD) software~\cite{andersonGeneralPurposeMolecular2008, pronkGROMACS45Highthroughput2013, thompsonLAMMPSFlexibleSimulation2022}. This limits their extensibility and practical usability in domain-specific workflows.

Moreover, the rapid pace of innovation in other aspects of the field presents additional challenges. New data curation strategies~\cite{zhangDPGENConcurrentLearning2020a, smithAutomatedDiscoveryRobust2021, levineOpenMolecules20252025}, training methodologies~\cite{thalerLearningNeuralNetwork2021, thalerDeepCoarsegrainedPotentials2022, rockenPredictingSolvationFree2024}, and schemes for incorporating long-range interactions~\cite{chengLatentEwaldSummation2025, fuchsLearningNonLocalMolecular2025a, kosmalaEwaldbasedLongRangeMessage2023, carusoExtendingRANGEGraph2025a, frankEuclideanFastAttention2024} are being developed continuously. Integrating these advancements into existing software frameworks often demands additional engineering effort, which can lead to a more diverse and specialized ecosystem. This fragmentation makes it harder for practitioners to adopt state-of-the-art methods and for developers to maintain robust, general-purpose tools.

Another emerging concern is how MLP performance is evaluated. While most efforts have focused on minimizing force and energy prediction errors, recent studies argue that simulation stability and the accuracy of observable quantities are more critical for practical applications~\cite{fu2022forces,fuLearningSmoothExpressive2025, potaThermalConductivityPredictions2024, loewUniversalMachineLearning2025, rajaStabilityAwareTrainingMachine2025}. Benchmarking different architectures against specific simulation tasks is thus gaining importance, as it provides insights that are more relevant for real-world usage and future model development.

In response to these challenges, several projects have attempted to provide easy-to-use interfaces between the ML core and different types of traditional modelling software.
Plugins have been developed such as Allegro-LAMMPS~\cite{musaelianAllegro2023}, SevenNet~\cite{parkScalableParallelAlgorithm2024}, GROMACS-NNpot~\cite{pronkGROMACS45Highthroughput2013}, FitSNAP~\cite{rohskopfFitSNAPAtomisticMachine2023}, and OpenMM-Torch~\cite{eastmanOpenMM8Molecular2023} to enable simulations with MD software such as LAMMPS~\cite{thompsonLAMMPSFlexibleSimulation2022}, GROMACS~\cite{pronkGROMACS45Highthroughput2013}, and OpenMM~\cite{eastmanOpenMM8Molecular2023}. However, many remain constrained by architecture-specific designs or face scalability challenges. DeepMD-kit~\cite{zengDeePMDkitV3MultipleBackend2025a} has recently introduced a multi-backend framework and support for external models~\cite{zengDeePMDGNNDeePMDkitPlugin2025a}, but its performance and scalability across multi-GPU systems have yet to be validated.

In this work, we present \texttt{chemtrain-deploy}, a model-agnostic deployment framework that extends our exsisting JAX-based training platform, \texttt{chemtrain}~\cite{fuchsChemtrain2025}. \texttt{chemtrain} was originally designed to support customizable training of neural network potentials with different training strategies, which integrates with JAX, M.D~\cite{schoenholzJAXMDFramework2020} to offer a unified training and simulation environment. However, current JAX, M.D. lack the robustness, interoperability, and scalability of established packages such as LAMMPS. \texttt{chemtrain-deploy} bridges this gap by enabling seamless deployment of pretrained semi-local MLPs into LAMMPS, allowing efficient large-scale simulations with systems containing millions of atoms across mutliple GPUs. To demonstrate flexibility and scalability, we benchmark \texttt{chemtrain-deploy} on three widely used GNN models: MACE, Allegro, and PaiNN, all trained to comparable accuracy. We test them on diverse systems including water–vapor coexistence, solid state (fcc) aluminum, and solvated Chignolin, evaluating strong and weak scaling, parallel efficiency, and performance relative to other frameworks such as JAX, M.D. Our results demonstrate the practicality of \texttt{chemtrain-deploy} for real-world, high-performance molecular simulation and provide guidance on GNN selection and future development.

\section{Results}

\paragraph{Structure of \ctd{}}

\begin{figure}[t]
    \centering
    \includegraphics[width=\linewidth]{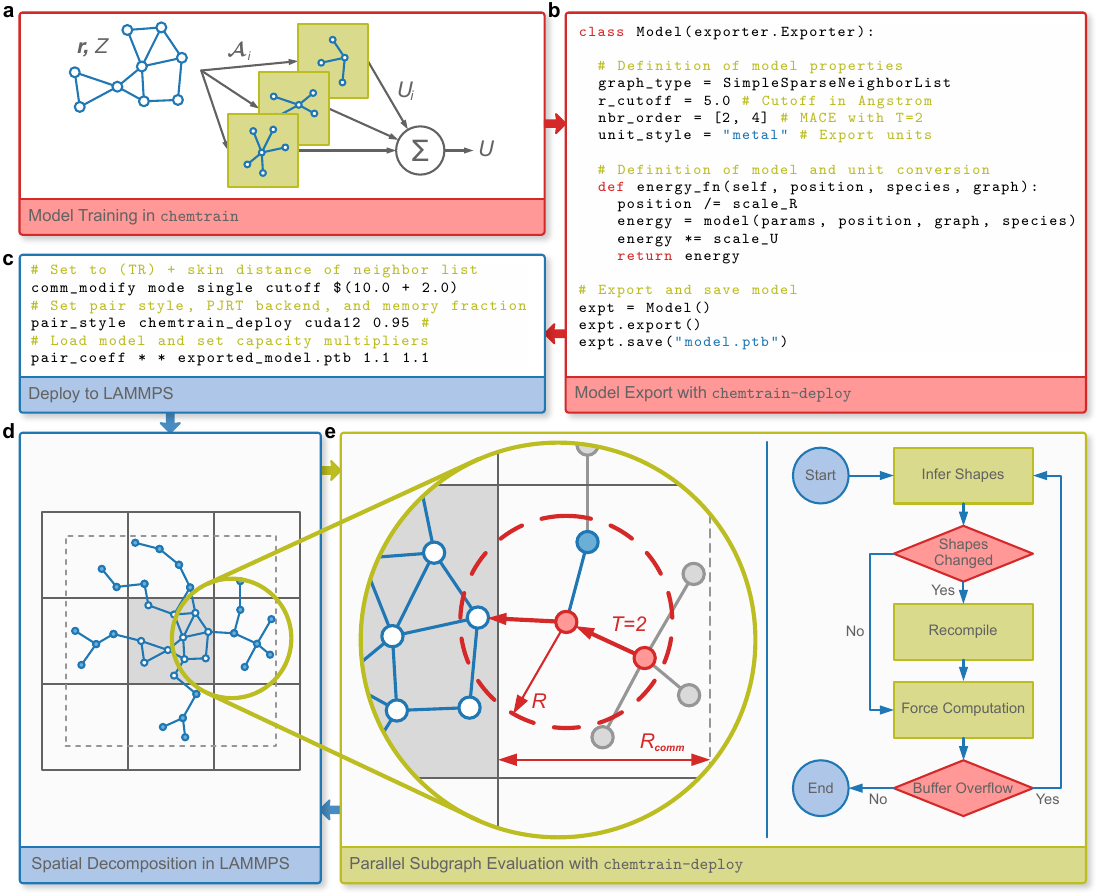}
    \caption{\textbf{Overview of \ctd{}.} Model trained in \texttt{chemtrain} \textbf{(a)} is exported \textbf{(b)} and loaded into LAMMPS \textbf{(c)}. LAMMPS distributes the workload onto multiple processors by decomposing the system into domains (solid lines) and computes domain neighbor lists including atoms in the domain (blue empty) and atoms from other domains (blue filled) within $R_\text{comm} \geq TR$ (dashed gray lines) \textbf{(d)}. \ctd{} computes the potential and forces independently for each domain, pruning atoms (gray) from the neighbor list graph generated by LAMMPS and buffering atom and graph data to fixed shapes required by XLA \textbf{(e)}.}
    \label{fig:chemtrain-deploy}
\end{figure}

\ctd{} complements the \verb|chemtrain|~\cite{fuchsChemtrain2025} framework to apply trained models in large-scale MD simulations using established MD software.
Therefore, \ctd{} comprises three parts: exporting a trained model, importing it into established MD software through a plugin or modification, and efficiently evaluating the model on high-performance hardware.
The parts and workflow of \ctd{} are depicted in Figure~\ref{fig:chemtrain-deploy}.

First, \ctd{} extends the framework \verb|chemtrain| (Fig. \sref{fig:chemtrain-deploy}{a}) to export trained models to a self-contained format (Fig.~\sref{fig:chemtrain-deploy}{b}). The export saves the model architecture and parameters through the MLIR framework~\cite{mlir} and properties that define the input and output of the model, such as length and energy units, the maximum length of graph edges, and the format of the input graph. Therefore, the exported model file contains all the information needed to apply the model and is thus simple to share or archive within the JAX compatibility guarantees.

Secondly, \ctd{} consists of a plugin to load and use the exported model for large-scale molecular dynamics simulations in established MD software (Fig.~\sref{fig:chemtrain-deploy}{c}).
The MD software provides basic and advanced algorithms to perform MD simulations.
Moreover, the MD software provides algorithms to decompose the system into multiple domains for parallelization and create a neighbor graph representation of the system (Fig.~\sref{fig:chemtrain-deploy}{d}).
Therefore, the plugin interfaces \ctd{} with framework-specific algorithms for parallelization and to run advanced MD simulations with the exported model.
In the current version, \ctd{} provides a plugin to the MD software LAMMPS~\cite{thompsonLAMMPSFlexibleSimulation2022}.

Finally, \ctd{} provides a library for the plugin to evaluate the exported model (Fig. \sref{fig:chemtrain-deploy}{e}).
This library uses XLA~\cite{xla2025github} and PJRT~\cite{PJRT} to translate the model into an efficient backend-specific computation at runtime.
Therefore, the library transforms and buffers the MD softwares's atom and neighbor data for compilation with XLA.
Following, the XLA compiler performs hardware-independent optimizations such as common subexpression elimination and operation fusion to reduce computational cost and memory requirements.
Finally, the pluggable PJRT runtimes further optimize the code for specific backends, such as GPUs and CPUs, considering the backend's architecture.
Thus, the library extends the LAMMPS' capabilities to run efficiently on specific hardware and evaluate new force-field architectures without rewriting or recompiling LAMMPS.
Moreover, the shared library promotes future extension of \ctd{} by reusing provided functionality in new plugins, respectively extensions, to other MD software.

\paragraph{Distributed potential computation}

In the following, we describe in more detail how \ctd{} approaches distributed potential and force computation for semi-local potentials (see section~\ref{subsec:semi-local-potential-models}).
For these potentials, we expect that the total potential energy of an $N$ atom system
\begin{align}
    U(\bm r) = \sum_{i=1}^N U_i\left(\bm r_i, \mathcal A_i\right)
\end{align}
decomposes into a sum of semi-local per-atom energies $U_i$ that depend on the position of the atom $\bm r_i$ and the atoms $\mathcal A_i$ within the semi-local environment of atom $i$.
Given a graph of the system, which represents atoms by nodes that share edges if closer to each other than a cutoff distance $R$, the local environment $\mathcal A_i = \left\{(\bm r_j, Z_j) \mid j \in \mathcal N_{\leq T}(i) \right\}$ contains positions $\bm r_j$ and species $Z_j$ of all atoms that are direct neighbors $\mathcal N_{=1}(i)$ of node $i$.
For semi-local potential models such as message-passing GNNs with $T$ message-passing steps, the local environment additionally contains atoms $i$, referred to as $T$-th order neighbors $\mathcal N_{\leq T}(i)$, to which a path of at most length $T$ exists (see Fig.~\sref{fig:chemtrain-deploy}{e}).

\ctd{} computes the potential energies in parallel on multiple independent processors (GPUs/CPUs) by partitioning the full graph into one subgraph per processor using spatial system decomposition and neighbor list generations provided in MD software, e.g., LAMMPS~\cite{thompsonLAMMPSFlexibleSimulation2022}.
As outlined in Figure~\sref{fig:chemtrain-deploy}{d}, LAMMPS distributes the workload by dividing the system into non-overlapping domains, such that each atom is local to exactly one domain.
LAMMPS then assigns the domains and the contained local atoms to the available processors.
For each processor, LAMMPS additionally copies all atoms from other domains within a distance of $R_\text{comm}$ of the processor's domain boundary and constructs a neighbor list graph.
Since the maximum distance between an atom $i$ and any of its $T$-th neighbors can be $TR$, choosing $R_\text{comm} \geq T R$ ensures that the neighbor graph of each domain contains all $T$-th neighbor atoms of all local atoms.
Thus, summing up the predicted energies of local atoms results in the total potential energy of the system, as each atom energy $U_i$ is computed exactly once on a subgraph with a complete environment of $i$.

Running MD simulations requires \ctd{} to compute the forces acting on the atoms.
From the sum rule, the total force on an atom
\begin{align}
    \bm f_i = -\frac{\partial U_i(\bm r_i, \mathcal R_i)}{\partial \bm r_i} - \sum_{j \in \mathcal N_{\leq T}^R} \frac{\partial U_j(\bm r_j, \mathcal R_j)}{\partial \bm r_i}
\end{align}
decomposes into a sum of partial forces $\bm f_{ij} = -\frac{\partial U_j(\bm r_j, \mathcal R_j)}{\partial \bm r_i}$.
The total forces of all local atoms can be computed directly on each processor by computing all nonzero partial forces.
However, this approach generally requires extending the domain subgraph to include the $2T$-th order neighbors, which are necessary to correctly compute the potential energy of all $T$-th order neighbors of the local atoms.
Alternatively, \ctd{} computes the partial forces of all local atoms with respect to all atoms of the domain subgraph.
Then, the total forces on each local atom can be obtained by summing up all partial forces from corresponding copies on other processors.
Thus, by constructing graphs containing all $T$-th order neighbors of local atoms, all particles' forces and potential energies can be computed with initial and final but without intermediate communication operations.

The graph obtained from LAMMPS might not be minimal and may contain copied atoms that are not within the semi-local environment of any local atom (gray nodes and edges in Fig.~\sref{fig:chemtrain-deploy}{e}).
Moreover, neighbor lists are typically constructed with edges longer than the model cutoff to prevent a costly neighborlist recomputation at every timestep.
Therefore, \ctd{} prunes the neighbor list graph at every timestep to improve the costly model evaluation.
First, \ctd{} removes all edges from the graph that are longer than the specified cutoff.
Following \ctd{} identifies all $T$-th neighbors of the local atoms by sending out pseudo-messages from the local atoms.
Each atom that received messages in the previous steps sends a message in the next step.
Finally, after the $T$ message passing steps, all $T$-th neighbors of local atoms have received a message.
\ctd{} automatically adds the pruning computation to the model during the export (Fig.~\sref{fig:chemtrain-deploy}{b}).
Therefore, pruning operations are parallelized and optimized through XLA.

\paragraph{Parallelization cost}
\label{subsec: parallel efficiency}

We estimate the cost of parallelizing semi-local potential models in homogeneous systems.
In homogeneous systems, the cost of a semi-local potential model scales linearly with the number of atoms in the domain~\cite{musaelianAllegro2023}.
We assume that the system domains are rectangular boxes with side lengths $L_x = L_y = L_z = L$ for bulk systems periodic in all dimensions and $L_x = L_y = L \gg L_z$ for surface systems periodic in the $x$ and $y$ dimension, the total number of atoms is $N$ is proportional to $L^d$, where $d$ is the number of periodic dimensions.
However, due to copied particles within a distance of $TR$ to the domain boundary, the cost of computing energies and forces is proportional to $(L + 2TR)^d$.

The workload can be divided among $P$ processors by using the domain decomposition described before to accelerate the computation.
Therefore, each processor computes forces and energies for a domain of approximately length $P^{-1/d}L$ in the periodic dimension, still requiring copies of atoms within $TR$ distance to the domain boundary.
Under the assumption that the runtime is proportional to the cost, the parallelization speeds up the computation by a factor
\begin{align}
    S = \left(\frac{L + 2TR}{P^{-1/d}L + 2TR}\right)^d \label{eq:approximate_strong}.
\end{align}
Since $P$ processors have to spend a relatively higher amount of work on computing interactions between copied atoms than between local atoms, the total work increases, causing a decrease in the parallel efficiency 
\begin{align}
    \varepsilon = \frac{S}{P}. \label{eq:approximate_efficiency}
\end{align}

\paragraph{Runtime optimizations and buffering}

\ctd{} optimizes and evaluates the model through XLA.
However, XLA re-optimizes the program every time the shape of an input changes, which typically requires more time than the actual computation.
Thus, \ctd{} buffers all dynamically shaped inputs to a fixed shape and evaluates the model as outlined in Figure~\ref{fig:chemtrain-deploy}.
First, \ctd{} computes the required buffer shapes. These shapes can vary, for example, if the number of atoms and neighbors in the domain changes.
If the buffer capacities are exceeded, \ctd{} enlarges the buffers and recompiles the model using the TensorFlow~\cite{tensorflow2025zenodo} call module loader.
If no buffer overflowed, \ctd{} transforms and copies data to the device and performs the computation.
The models use internal buffers to enable optimizations such as graph pruning in the computation.
Thus, after the computation, \ctd{} checks whether internal model buffers overflowed.
If internal model buffers overflow, \ctd{} repeats the computation with resized model buffers.
If no buffer overflowed, \ctd{} copies back the computed forces and returns statistics of the computation.

Since \ctd{} only enlarges buffers, the frequency of recompilations decreases for systems at equilibrium.
However, recompilations can happen frequently in the initial stages of computations on multiple devices if each device recompiles independently.
Thus, \ctd{} enforces collective recompilations of multiple devices per time step by explicitly controlling recompilations.
Therefore, the \ctd{} plugin first tries to evaluate the model with recompilation disabled.
If a recompilation is necessary on one device, the device raises an exception that will be called by the plugin.
The plugin then synchronizes the error to all devices, which will enlarge overflown and nearly filled buffers and recompile the program.
Thereby, \ctd{} boosts simultaneous recompilations on multiple devices to shorten warm-up periods in large-scale parallel applications.

\paragraph{Traning state-of-the-art neural MLPs with \texttt{chemtrain}}
With the flexibility of \texttt{chemtrain} in training state-of-the-art MLPs, we used it to train models on three chemically and structurally diverse systems commonly studied in biophysics and materials science: a liquid-vapor water system, a crystalline aluminum solid, and the mini-protein Chignolin solvated in water. These systems span homogeneous and heterogeneous environments and include different phase states, such as liquid, solid, and interfacial configurations, capturing a broad range of chemical and structural complexity. For each system, we chose a corresponding training dataset: H2O-PBE0TS~\cite{zhangDeepPotentialMolecular2018a}, ANI-AL~\cite{smithAutomatedDiscoveryRobust2021}, and SPICE~\cite{eastmanSPICEDatasetDruglike2023}, respectively. We used three different GNN architectures that reflect the methodological diversity of modern approaches: Allegro~\cite{musaelianAllegro2023}, MACE~\cite{batatiaMACEHigherOrder2023}, and PaiNN~\cite{schuttEquivariantMessagePassing2021}, on the same dataset for each system. For fair comparison, we carefully selected hyperparameters to achieve similar energy and force accuracy across architectures. In all cases, the models reached comparable MAE or RMSE values, remained within chemical accuracy, and matched reported literature benchmarks (Table~\ref{table: model_accuracy}). Further details on the datasets, training procedures, and hyperparameter configurations are provided in the Methods section.

\begin{table}[htb]
\centering
\caption{Root mean square (RMSE) and mean absolute errors (MAE) for energies (meV/atom) and forces (meV/Å) across Allegro, MACE, and PaiNN models on the ANI-AL, SPICE, and H\textsubscript{2}O-PBE0TS datasets, with reference values from the literature, including classical MEAM and other MLPs.}
\label{table: model_accuracy}
\begin{tabular}{lcccc}
\toprule
 & \textbf{Allegro} & \textbf{MACE} & \textbf{PaiNN} & \textbf{Reference} \\
\midrule
\textbf{ANI-AL} & & & & ANI-AL~\cite{smithAutomatedDiscoveryRobust2021}, MEAM~\cite{leeMEAMAluminium2003}\\
\hskip 1em Energy (RMSE) & 12.9 & 9.9 & 7.2 & 1.9, 60.6 \\
\hskip 1em Force (RMSE)  & 109.4 & 71.8 & 62.7 & 60.0, 244.8 \\
\midrule
\textbf{SPICE} & & & &  TorchMD-NET~\cite{eastmanSPICEDatasetDruglike2023}\\
\hskip 1em Energy (MAE)  & 12.4 & 46.1 & 27.4 & 48.3 \\
\hskip 1em Forces (MAE)  & 73.5 & 47.6 & 48.4 & -- \\
\midrule
\textbf{H\textsubscript{2}O-PBE0TS} & & & & NequIP~\cite{batznerNequIP2022}, DeepMD~\cite{zhangDeepPotentialMolecular2018a}\\
\hskip 1em Energy (RMSE) & 0.7 & 0.5 & 0.7 & 0.6,  0.3 \\
\hskip 1em Force (RMSE)  & 36.2 & 13.3 & 17.2 & 11.6, 40.4 \\
\bottomrule
\end{tabular}
\end{table}

\paragraph{Memory requirements}


GNN-based MLPs can require significant memory to store high-dimensional node features and messages.
However, model-agnostic software such as JAX, M.D., GROMACS-NNPot, or OpenMM-Torch does not support multi-GPU simulations through domain decomposition.
Therefore, we determined the maximally supported system sizes and runtimes for MD simulations using GNN potentials that can be run on a single GPU with JAX, M.D.
Additionally, we report reference measurements for \ctd{}, which, unlike JAX, M.D., is not limited to only one GPU.

We tested all combinations of models and systems, for which we report accuracies in the previous section.
To increase the system sizes, we replicated all systems equally in all periodic dimensions as described in Section~\ref{subsec:md_simulations}.
For JAX, M.D., the maximum system sizes were lower than for \ctd{} (Table~\ref{table:jaxmd_comparison}), limited to less than half a million atoms. 
In comparison, \ctd{} could simulate more or a similar number of atoms than JAX, M.D. with the strictly local Allegro model on one GPU.
Differently, for the message-passing models, the maximum system sizes that could be simulated with \ctd{} were lower than for JAX, M.D.
In all cases, the runtime \ctd{} was similar or slower than for JAX, M.D. (Supplementary~Table~1).

The difference in memory consumption and computational efficiency is likely due to JAX, M.D. updating the neighbor list entirely on the GPU and applying periodic boundary conditions without copied atoms.
For large systems and short-ranged models, the neighbor list generation can require more memory in JAX, M.D. than computing interactions for copied atoms in \ctd{}.
For smaller systems and models with larger effective cutoffs, memory and compute requirements for copied atoms are higher than for local atoms, affecting the computational efficiency (see Supplementary~Figure~1).
However, due to the copied atoms, \ctd{} can parallelize the simulation on multiple GPUs.
Therefore, using additional GPUs could compensate for the higher memory requirements.
In contrast, JAX, M.D does not support parallelization, such that the memory requirements limit the maximum system sizes below the order of a million atoms.
However, applications such as the investigation of solidification can still exhibit finite-size effects up to two million atoms~\cite{Mahata2019FiniteSizeEffects}.
Thus, the single-GPU support prevents software such as JAX,~M.D. from deploying GNN potentials to applications requiring large-scale simulations.

\begin{table}[htb]
\centering
\caption{Maximum system sizes in number of atoms for JAX, M.D. vs \texttt{chemtrain-deploy} on a single GPU (A100, 80GB) for Allegro, MACE, and PaiNN applied to solid state aluminium (fcc) at 1000 K, replicated box of solvated Chignolin at ambient conditions, and water slab at ambient conditions.}
\label{table:jaxmd_comparison}
\begin{tabular}{@{}llcrr}
\toprule
 & \textbf{System}  & \textbf{JAX, M.D.} & \textbf{\ctd} \\ \midrule
\multirow{3}{*}{Allegro}  & Aluminium  & 296,352 & 470,596 \\
                          & Chignolin  & 27,936 & 27,936  \\
                          & Water      & 253,125 & 496,125 \\\midrule
\multirow{3}{*}{MACE}     & Aluminium  & 202,612 & 108,000 \\
                          & Chignolin  & 94,284 & 94,284 \\
                          & Water      & 162,000 & 112,500 \\\midrule
\multirow{3}{*}{PaiNN}    & Aluminium  & 340,736 & 171,500 \\
                          & Chignolin  & 27,936 & 3,492 \\
                          & Water      & 72,000 & 40,500 \\
\bottomrule
\end{tabular}
\end{table}

\paragraph{Scaling to million-atom systems}

\begin{figure}
    \centering
    \includegraphics[width=\textwidth]{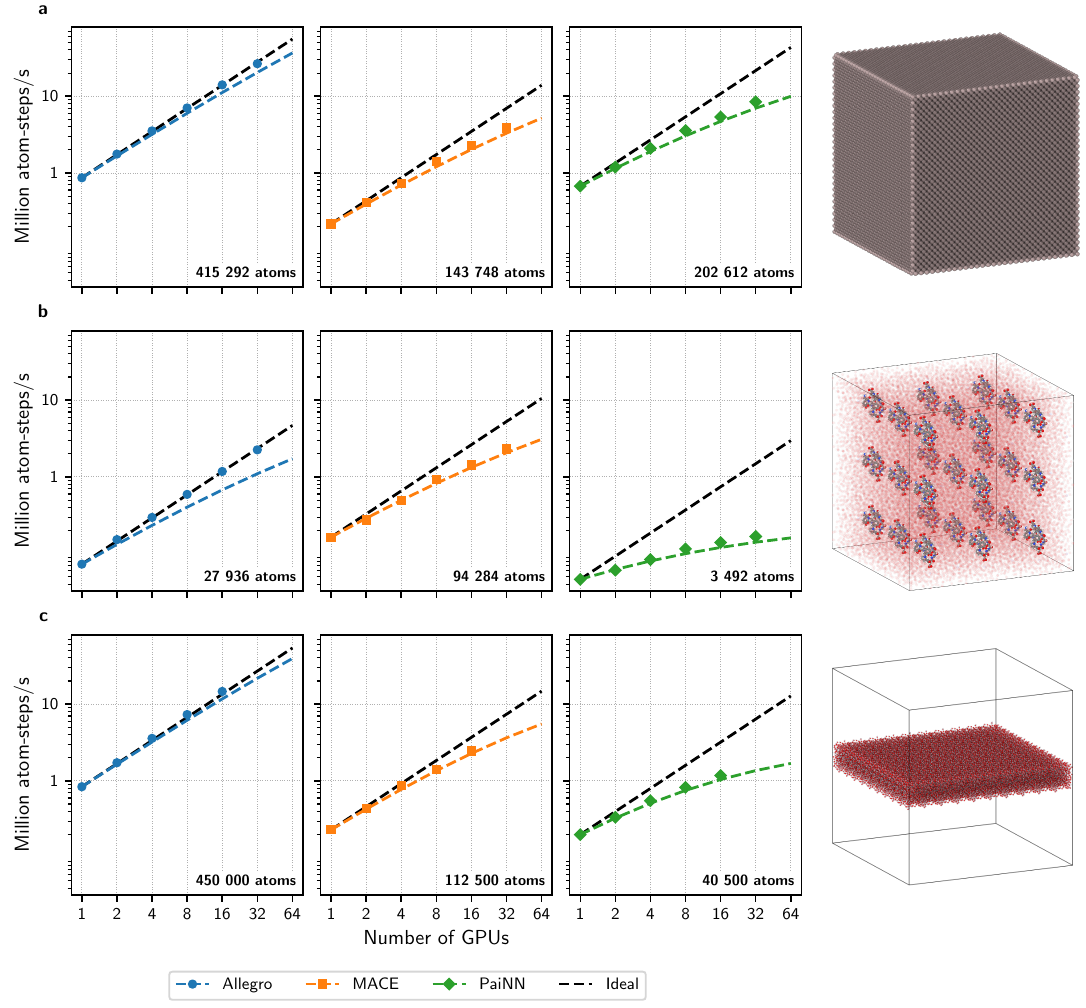}
    \caption{Strong scaling on JEDI for Allegro (blue), MACE (orange), and PaiNN (green) applied to \textbf{a} solid state aluminium (fcc) at 1000 K, \textbf{b} replicated box of solvated Chignolin at ambient conditions, and \textbf{c} water slab at ambient conditions next to visualizations of the systems. The sizes of each system are given in numbers of atoms in the lower right corner. Visualizations were created with OVITO~\cite{ovito} for the systems corresponding to the MACE model. Ideal and approximate (Eq.~\ref{eq:approximate_strong}) strong scaling are shown as dashed lines in black and the model colors, respectively. }
    \label{fig:strong_scaling}
\end{figure}

To estimate the performance of \ctd{} for simulating large systems on multiple GPUs, we evaluated strong and weak scaling for all combinations of systems and models.
For each combination, we selected a different system size to respect the different memory requirements of the models.
For the strictly local Allegro model, we observed close-to-ideal strong scaling (Figure~\ref{fig:strong_scaling}), slightly outperforming the anticipated strong scaling in Eq.~\ref{eq:approximate_strong} and often exceeding the anticipated ideal parallel efficiency (Supplementary~Figure~2).
This improvement might be due to XLA optimizations leveraging additional memory and compute resources.
For the message-passing GNNs, we obtained good strong scaling, except for Chignolin simulated with the PaiNN model.
In all cases, the measured strong scaling is consistent with our approximation given in Eq.~\ref{eq:approximate_strong} for all systems.

\begin{figure}
    \centering
    \includegraphics[width=0.35\linewidth]{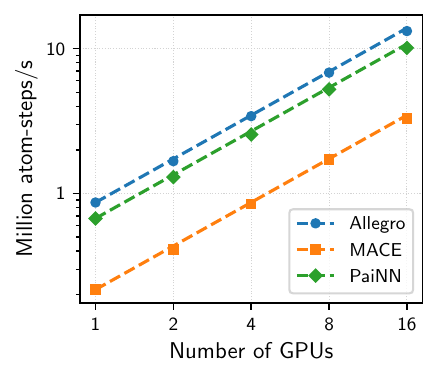}
    \caption{Weak scaling on JEDI for Allegro (blue), MACE (orange), and PaiNN (green) applied to solid state aluminium (fcc) systems at 1000 K with $415,292$, $143,748$, and $202,612$ atoms per GPU.
    Ideal weak scaling is displayed as dashed lines for each model in the respective color.}
    \label{fig:weak_scaling}
\end{figure}

As shown in Figure~\ref{fig:weak_scaling}, all models exhibited close-to-ideal weak scaling.
This result indicates that inter-device and
inter-node communications do not crucially affect efficiency for multi-GPU computations.
Therefore, scaling in \ctd{} is mostly determined by the effort spent on copied atoms compared to local atoms (Supplementary~Figure~1).
Thus, Eq.~\ref{eq:approximate_strong} provides a good reference to estimate the required cost and resources of scaling message-passing GNNs to large systems.

To directly compare the models, we also evaluated the scaling on million-atom systems, visualized in Figure~\ref{fig:total_scaling} (simulation speeds reported in Supplementary Table~2).
The Allegro and MACE models showed good strong scaling for all systems.
The PaiNN model scaled slightly worse than the other models in the aluminium system and failed for the Chignolin and water systems due to insufficient memory.
Comparing the simulation speed of all models, the Allegro model performed best for aluminium and water.
The MACE models achieved a similar throughput for all systems, outperforming the speed of Allegro for the Chignolin system.
The PaiNN model performed similarly to the Allegro model in the aluminium system.
Based on these scaling results, we conclude that the Allegro model is highly efficient for systems with a few atom types, such as water and aluminium.
However, for chemically diverse systems, MACE provides a better tradeoff between accuracy, scalability, and robustness (Supplementary Note~1).
Using PaiNN with a larger effective cutoff can be beneficial for very simple systems, but requires extensive memory for chemically more diverse systems.

We additionally compare the scaling with other implementations and models as a reference.
For the aluminium system, all models scaled better than a classical MEAM potential~\cite{leeMEAMAluminium2003} (Figure~\ref{fig:total_scaling}) but were slower for the same number of GPUs.
The computational speed of the MEAM potential is naturally higher due to a simpler computation, which affects the model's accuracy (Table~\ref{table: model_accuracy}).
The Allegro models for the aluminium and water system showed strong scaling comparable to Allegro models deployed to similar systems with Allegro-LAMMPS~\cite{musaelianAllegro2023,kozinskyScalingAllegro2023} (Supplementary~Figure~3). 
However, the exact difference in computational speed depends on the model architecture, such as the depth of the model and the cutoff of the graph.

\begin{figure}
    \centering
    \includegraphics[width=.75\linewidth]{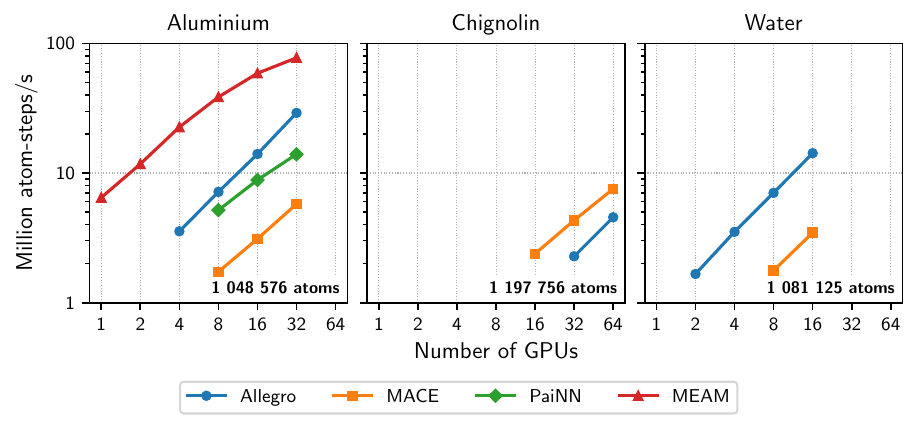}
    \caption{Strong scaling on JEDI for Allegro (blue markers and lines), MACE (orange markers and lines), and PaiNN (green markers and lines) applied to solid state aluminium (fcc) at 1000 K (left), replicated box of solvated Chignolin at ambient conditions (middle), and water slab at ambient conditions (right) for systems with approximately 1 million atoms. The exact numbers of atoms are shown in bold in the lower right corners of the plots. Missing results correspond to simulations that failed due to insufficient memory. Scaling for the modified embedded atom method (MEAM) potential~\cite{leeMEAMAluminium2003} applied to the aluminium system is shown as reference (red markers and line).}
    \label{fig:total_scaling}
\end{figure}

\section{Discussion}

We present \texttt{chemtrain-deploy}, a model-agnostic framework for deploying JAX-based semi-local MLPs to LAMMPS. By coupling JAX-based models with the scalability and functionality of LAMMPS, \texttt{chemtrain-deploy} provides a seamless interface for running complex and large-scale simulations with minimal integration overhead on multiple GPUs. Therefore, \ctd{} overcomes key limitations of existing software, which are often restricted to specific model architectures, provide limited training support, or face scalability challenges.

Our results demonstrate excellent scaling of modern GNN potentials through \ctd{} for different systems, particularly in simulations involving millions of atoms across multiple GPUs. Through optimizations such as XLA-based compilation and graph pruning, we reduce execution overhead and minimize recompilation costs. This capability enables the application of semi-local MLPs to new fields in computational biology and material sciences. 

We compared different state-of-the-art GNN architectures. We found that strictly local models generally exhibit superior scalability due to their limited communication overhead, while semi-local message-passing GNNs tend to provide improved accuracy but can exhibit reduced scalability and increased memory demands. Nonetheless, actual computational performance depends on the specific MLP hyperparameters and the systems of interest to practitioners. These insights might provide a starting point for future users to select architectures and guide the development of MLPs.

\texttt{chemtrain-deploy} can also support other semi-local MLPs beyond those models demonstrated in this paper, such as Behler-Parinello potentials~\cite{behlerGeneralizedNeuralNetworkRepresentation2007}, NequIP~\cite{batznerNequIP2022}, DimeNet++~\cite{gasteigerFastUncertaintyAwareDirectional2022}, and future MLPs likely to be available in a JAX-based implementation. Moreover, these architectures can be combined, e.g., with classical field priors for coarse-grained systems~\cite{wangMachineLearningCoarseGrained2019, thalerLearningNeuralNetwork2021}, or Coulomb interactions and dispersion corrections~\cite{kabyldaMolecularSimulationsPretrained2024c} to form hybrid classical/GNN potentials with high stability and effective treatment of long-range interactions~\cite{chengLatentEwaldSummation2025, fuchsLearningNonLocalMolecular2025a, kosmalaEwaldbasedLongRangeMessage2023, carusoExtendingRANGEGraph2025a, frankEuclideanFastAttention2024}.
From a future perspective, the flexibility and extensibility of \texttt{chemtrain-deploy} ensure its long-term usability beyond current state-of-the-art models and foster innovation through enabling rapid implementation and testing.

Looking forward, several promising avenues exist to extend the capabilities of \texttt{chemtrain-deploy}. These include support for global models that can capture long-range interactions~\cite{fuchs2025learningnonlocalmolecularinteractions}, integration with other popular MD software such as GROMACS~\cite{pronkGROMACS45Highthroughput2013} and NAMD~\cite{phillips2020scalable}, implementation of adaptive cutoff schemes~\cite{kozinskyScalingAllegro2023}, and multiscale modeling techniques such as multi-time-step algorithms~\cite{tuckermanReversibleMultipleTime1992}. Such developments will expand the applicability of \texttt{chemtrain-deploy} to an expanded set of complex systems and simulation scenarios, further bridging the gap between MLPs and practical, large-scale molecular simulations.

The broad applicability of \texttt{chemtrain-deploy} opens several exciting opportunities for MD community. Its ability to efficiently handle million-atom systems enables simulations of complex materials and biological systems. By making use of the various functions of LAMMPS, \texttt{chemtrain-deploy} supports non-standard simulation protocols such as those under external fields or using enhanced sampling techniques, expanding the range of molecular phenomena that can be studied. Importantly, its model-agnostic design allows easy benchmarking and comparison of different machine learning potentials within consistent simulation settings, accelerating the development and validation of next-generation models. Together, we expect that \texttt{chemtrain-deploy} will accelerate the adoption and development of advanced machine learning potentials, enabling transformative advances in large-scale molecular simulations and ultimately driving progress across computational chemistry and materials science.

\section{Methods}
\subsection{(Semi-)Local Potential Models}
\label{subsec:semi-local-potential-models}

Molecular dynamics simulations can describe the behavior of atomistic systems through forces $\bm f_i$ that derive from a potential energy function $\bm f_i = \frac{\partial U(\bm r)}{\partial \bm r_i}$.
In many systems, atoms predominantly interact with other atoms in their local environment.
Therefore, many classical and modern models approximate the total energy of a system through a sum of local atomic energy contributions
\begin{align}
    U(\bm r) = \sum_{i=1}^N U_i\left(\bm r_i, \mathcal A_i\right),
\end{align}
where $\mathcal A  = \left\{\left(\bm r_j, Z_j\right) \mid j \in \mathcal N_{=1}(i) \right\}$ is the set of atom positions and species $Z_i$ of the direct neighbors $\mathcal N_{=1}(i)$ to particle $i$ with a distance $\lVert\bm r_{ij}\rVert$ less than a cutoff.~\cite{behlerHDNNP2007,musilPhysicsInspiredStructuralRepresentations2021}

\paragraph{Descriptors}
The predicted potential energy should be invariant to permutations of atoms of the same species and translation, rotation, and reflections of the reference frame~\cite{behlerHDNNP2007, musilPhysicsInspiredStructuralRepresentations2021}.
However, applying general regression models such as Neural Networks or Kernel Models directly to the atomic positions does not necessarily result in a model that respects these invariances~\cite{behlerHDNNP2007}.
Therefore, many potential models first encode the atomic environments into an invariant vector of local descriptors $\bm \xi = \left\{\xi^\alpha\left(\bm r_i, \mathcal A_i \right)\right\}$ that serve as input to an learnable atomic energy function $U_i(\bm r, \mathcal A_i) = \tilde U_i\left(\bm \xi_i\left(\bm r_i, \mathcal A_i\right)\right)$~\cite{musilPhysicsInspiredStructuralRepresentations2021} such as a Neural Network~\cite{behlerHDNNP2007}.
Using the same model and descriptors for all particles of the same species ensures geometric and permutation invariant potential predictions~\cite{behlerHDNNP2007}.

Local descriptors typically achieve translational invariance by acting on the atom displacements $\bm r_{ij}$ rather than the absolute positions~\cite{musilPhysicsInspiredStructuralRepresentations2021}.
Moreover, descriptors such as the Atom-Centered Symmetry Functions (ACSF) achieve rotation and reflection invariance by encoding distances and directions between pairs and triplets of neighbors through invariant distances and angles.
However, this encoding scales unfavorably with the number of neighbors for more than two-body correlations.
Therefore, equivariant descriptors such as the Smooth Overlap of Atomic Positions (SOAP)~\cite{bartokRepresentingChemicalEnvironments2013} represent displacements in a suitable basis and perform equivariant operations to encode correlations between multiple neighbors while scaling linearly with the number of neighbors.
More generally, the Atomic Cluster Expansion (ACE) provides a systematic approach to constructing a complete basis for the local environment through a hierarchical expansion. Thus, the ACE descriptor can represent many previously proposed local descriptors while scaling linearly with the number of neighbors~\cite{drautzAtomicClusterExpansion2019}.

\paragraph{Graph Neural Networks}
Devising accurate and efficient descriptors by hand for chemically diverse datasets is difficult.
Approaches such as ACE describe how to systematically build a complete basis for the local environment~\cite{musilPhysicsInspiredStructuralRepresentations2021}.
However, the resulting descriptors scale unfavorably with the number of atom species due to the number of basis functions~\cite{musaelianAllegro2023}.
Thus, learning efficient environment descriptors from data through Graph Neural Networks (GNNs) has gained wide attention.

GNNs encode locality by representing the system as a graph, with nodes representing atoms and edges connecting all neighbors $\mathcal N_{=1}$.
The graph is commonly embedded by assigning node features $h^{(0)}_i = f_h(Z_i)$ based only on the particle species to ensure permutation invariance, and edges features $e^{(0)}_{ij} = f_e(\bm r_{ij})$ based on edge displacements to ensure translational invariance.
Many proposed models, such as SchNet~\cite{schuttSchNet2018} or DimeNet~\cite{gasteigerDirectionalMessagePassing2022}, then extract environment descriptors through the Message-Passing (MP) framework~\cite{gilmerMessagePassingGNN2017} by propagating information along the graph through messages
\begin{align}
    \bm m_i^{t+1} & = \sum_{j\in \mathcal N(i)} \mathcal M^{t}(\bm h_i^t, \bm h_j^t, \bm e_{ij}),
\end{align}
which aggregate information from neighboring atoms by a learnable message function $\mathcal M^t$.
Using the aggregated messages, the MP-GNNs then update the node features
\begin{align}
    \bm h_i^{t+1} & = \mathcal U^{t}(\bm m_i^{t+1}, \bm h_i^t),
\end{align}
through an update function $\mathcal{U}^t$.
The final node features $\bm h_i^L$ after $T$ message passing steps can act as input to a regression model such as a Neural Network~\cite{schuttSchNet2018} or a Gaussian Process~\cite{wollschläger2023uncertaintyestimationmoleculesdesiderata}.

Similar to classical descriptors, the GNN predictions must be invariant to translations, rotations, and reflections.
Early GNNs achieved this invariance through an invariant graph embedding using distances~\cite{schuttSchNet2018} and angles~\cite{gasteigerDirectionalMessagePassing2022}.
However, higher-order embeddings can improve the GNN expressiveness but scale unfavorably with the number of neighbors~\cite{musaelianAllegro2023}, similar to ACSFs.
Moreover, propagated messages contain only invariant information about the particle's local environment, prohibiting leveraging information about their relative orientation~\cite{schuttEquivariantMessagePassing2021}.
Therefore, equivariant MP-GNNs generalize message-passing to tensorial features, such as displacements between atoms, to efficiently propagate directional information about atomic environments.
Thereby, equivariant GNNs employ operations that ensure tensorial features are equivariant, i.e., transform similarly to the input for a group of transformations, to ensure invariance of the final scalar node features~\cite{batznerNequIP2022,schuttEquivariantMessagePassing2021,batatiaDesignSpaceE3equivariant2025}.

Unlike the previously described strictly local descriptors, messages passing GNNs can propagate information between atoms that are not direct neighbors.
Instead, GNNs propagate information from an atom $i$ to higher-order neighbor atoms $\mathcal N_{\leq T}$ connected by a path of length less or equal than $T$.
Thus, MP-GNNs with many message-passing layers have a large receptive field of radius $TR$, potentially impairing their scalability.
To ensure high scalability, multiple approaches aim at constructing descriptive GNNs with small receptive fields.
The Multi-ACE framework~\cite{batatiaDesignSpaceE3equivariant2025} reformulates message construction in the ACE formalism, generalizing previous invariant and equivariant MP-GNNs, such as SchNet~\cite{schuttSchNet2018}, DimeNet~\cite{gasteigerDirectionalMessagePassing2022}, NequIP~\cite{batznerNequIP2022}, and PaiNN~\cite{schuttEquivariantMessagePassing2021}, that correlate only information from a limited number of neighbors for each message.
Through the ACE formalism, this framework enables models such as MACE~\cite{batatiaMACEHigherOrder2023} to construct messages that correlate information from an arbitrary number of neighbors to exploit high-order many-body correlations independently of the number of message-passing steps while scaling linearly with the number of neighbors.
The Allegro model~\cite{musaelianAllegro2023} reformulates message-passing in an edge-centric formalism.
Therefore, Allegro only passes messages between directed edges originating from the same node. Consequently, no information is propagated to particles outside the cutoff shell. Consequently, the Allegro model learns strictly local environment descriptions.

\paragraph{Chosen Architectures}

In this work, we chose the GNN models PaiNN, MACE, and Allegro as examples of different design choices of models in terms of receptive fields and fidelity of semi-local descriptions.
On the one hand, the PaiNN model only correlates pairs of neighbor features in each message-passing layer, such that the number of message-passing layers determines the receptive field and the body order of the final descriptor.
On the other hand, the Allegro model employs equivariant edge-based message passing to ensure strict locality, such that the number of message-passing layers only affects the body order but not the receptive field of the model.
The MACE model employs the ACE formalism to correlate information from a variable number of neighbors.
Therefore, the final body order can be enlarged without increasing the receptive field.

\subsection{Reference Training Datasets}
\paragraph{ANI-AL}
The dataset includes over 6000 DFT calculations on supercells containing up to 250 atoms, covering a wide range of nonequilibrium configurations. It was generated using a minimally guided active learning approach. The data can be obtained from \url{https://github.com/atomistic-ml/ani-al}.

\paragraph{SPICE}
SPICE is a quantum chemistry dataset for simulating drug-like small molecules and proteins. It contains over 1.1 million configurations, representing a diverse set of small molecules, dimers, dipeptides, and solvated amino acids. The dataset provides energies and forces computed using the $\omega$B97M-D3(BJ)/def2-TZVPPD level of theory. The data can be obtained from \url{https://github.com/openmm/spice-dataset}.

\paragraph{H\textsubscript{2}O-PBE0TS}
The H\textsubscript{2}O-PBE0TS dataset contains snapshots of liquid water and ice configurations generated via ab initio molecular dynamics (AIMD) using the PBE0+TS functional. The data can be obtained at \url{https://aissquare.com/datasets/detail?pageType=datasets&name=H2O-PBE0TS}.

\subsection{Molecular Dynamics Simulations}
\label{subsec:md_simulations}

MD simulations were performed using \textsc{LAMMPS} with the \texttt{chemtrain-deploy} pair style implemented in our custom \texttt{chemtrain-deploy} interface. All simulations, including production and timing runs, were carried out in the NVT ensemble using a Nosé–Hoover thermostat with a temperature damping parameter of 1.1 ps. Each system underwent 100 equilibration steps followed by 250 production steps. For the aluminium case, a face-centered cubic lattice (lattice constant 4.065~\AA) was constructed and replicated equally in all three dimensions; simulations were conducted at 1000 K with a 3 fs time step and a 2.0~\AA{} neighbor skin. For the Chignolin case, the system was read from a preconfigured structure created by GROMACS, consisting of a Chignolin molecule solvated in a 3.3 nm cubic TIP3P water box. The system was then replicated equally in all three spatial dimensions according to the scaling index. Simulations were performed at 293.15 K using a 0.5 ps time step and a 2.5~\AA{} neighbor skin distance. For the water–vapor interface case, the system was initialized from a pre-equilibrated 2$\times$2$\times$5 nm$^3$ TIP3P water box and replicated equally in the $x$ and $y$ directions according to a scaling index; the $z$-direction was extended to create vacuum regions for a water–vacuum interface, similar to prior setup~\cite{sanchez-burgosDeepPotentialModel2023}. This simulation was run at 293.15 K using a 1 fs time step and a 2.5~\AA{} neighbor skin distance.  Simulations performed using JAX, M.D. followed the same settings and initial configurations as the corresponding \textsc{LAMMPS} simulations.

Scaling simulations were conducted on the JEDI test system using up to 16 nodes interconnected via InfiniBand NDR200. Each node consists of 4 NVIDIA GH200 Superchips, with each Superchip pairing 72 CPU cores and an H100 GPU with 96GB of memory. JAX-M.D. simulations were performed on a single A100 GPU with 80GB of memory.

\subsection{Training Details}

The following training settings are kept consistent across all models, with architecture-specific hyperparameters detailed in the corresponding subsections. All models are trained using a force-matching approach. Given a dataset of atomic configurations with reference energies \( U^{\text{ref}} \) and reference forces \( \bm f^{\text{ref}} \), we optimize the neural network parameters \( \theta \) to minimize differences between predicted and reference values. The training loss is defined as:
\begin{equation}
\mathcal{L}(\theta) = \lambda_E \sum_{\alpha} \left| U^\theta(\bm r_\alpha) - U^{\text{ref}}(\bm r_\alpha) \right|^2 + \lambda_F \sum_{\alpha, i} \left\| \bm f_i^\theta(\bm r_\alpha) - \bm f_i^{\text{ref}}(\bm r_\alpha) \right\|^2,
\end{equation}
where \( \lambda_E \) and \( \lambda_F \) control the relative weighting of the energy and force terms. All models are trained using single-precision (FP32) arithmetic and the Adam optimizer with default parameters: \( \beta_1 = 0.9 \), \( \beta_2 = 0.999 \), and \( \epsilon = 10^{-8} \). For each dataset, a consistent train–test–validation split ratio of 7:2:1 is used, where we used the validation split to select the best performing parameters. For MACE, the energy and force weights in the loss function are set to \( \lambda_U = 10^{-6} \) and \( \lambda_F = 10^{-1} \), respectively. For PaiNN, we use \( \lambda_U = 10^{-4} \) and \( \lambda_F = 10^{-1} \). For Allegro on water and aluminum, the weights are \( \lambda_U = 10^{-6} \) and \( \lambda_F = 10^{-1} \), while for Chignolin, both are set to \( \lambda_U = 10^{-4} \) and \( \lambda_F = 10^{-4} \).

All models use a graph cutoff distance of 5~\AA{}. For all Allegro models on water and alumnium, we use one tensor product layer with $l_{\text{max}} = 3$, 8 radial basis functions, and a polynomial envelope of order 2, while we use three layers with $l_{\text{max}} = 2$ for Chignolin. For MACE models, we set the hidden irreducible representations to $"32\times0e + 32\times1o"$ across all cases, with $l_{\text{max}} = 3$, a correlation order of 3 per layer, 2 interaction layers, a node embedding dimension of 64, 8 radial basis functions, and a polynomial envelope of order 6. For PaiNN, we use 4 layers in all cases and vary only the size of the embedding features, keeping all other hyperparameters fixed. The learning rate generally follows a polynomial decay schedule with a power of 2.0 and a decay rate  \( 10^{-5} \). The only exception is for the Allegro model on the SPICE dataset, where we use an exponential decay schedule with a decay rate of 0.001.

All models were trained on a single NVIDIA A100 GPU.

\paragraph{H\textsubscript{2}O-PBE0TS and ANI-AL Models}
The H\textsubscript{2}O-PBE0TS models were trained on a total of 100{,}000 samples. For Allegro, we use a hidden MLP layer dimension of 64 and embedding dimensions of [8, 16, 32]. The hidden irreducible representations are set to $"32\times0e + 16\times1e + 16\times1o + 8\times2e + 8\times2o"$. For both MACE and Allegro, the learning rate is set to 0.01. For PaiNN, we use a hidden feature size of 128 with a initial learning rate of 0.001.

The ANI-AL models were trained on 6{,}000 samples. All hyperparameters are kept the same as in the H\textsubscript{2}O-PBE0TS case, except that the PaiNN hidden feature size is set to 64. The learning rates remain the same, while the batch sizes are 64 for Allegro, 16 for MACE, and 8 for PaiNN.

\paragraph{SPICE Models}
We use the entire dataset for training, excluding the Ion Pairs subset, with a total of 1{,}817{,}199 samples. For Allegro, the hidden MLP layer dimension is set to 256, with embedding dimensions of [128, 128, 256]. The hidden irreducible representations are set to $"64\times1o + 16\times2e"$. For both Allegro and MACE, the initial learning rate is 0.001, with batch sizes of 16. For PaiNN, we use a hidden feature size of 128, a learning rate of $10^{-4}$, and a batch size of 32.

\section*{Acknowledgments}
Funded by the European Union. Views and opinions expressed are however
those of the author(s) only and do not necessarily reflect those of the European Union or the
European Research Council Executive Agency. Neither the European Union nor the granting
authority can be held responsible for them.
This work was funded by the ERC (StG SupraModel) - 101077842 and the Deutsche Forschungsgemeinschaft (DFG, German Research Foundation) - 534045056 and 561190767.

We gratefully acknowledge the Gauss Centre for Supercomputing e.V. (www.gauss-centre.eu) for supporting this project with computing time provided through the John von Neumann Institute for Computing (NIC) on the GCS Supercomputer JEDI at the Jülich Supercomputing Centre (JSC).

The authors thank Jan Eckwert and Ian Störmer for valuable discussions and Mario Geiger for open-sourcing his Allegro JAX code.

\section*{Data Availability}
The H\textsubscript{2}O-PBE0TS, ANI-AL and SPICE datasets are publicly accessible. (see “Methods”)
The parameters for the MEAM potential for aluminium can be accessed at \url{https://www.ctcms.nist.gov/potentials/entry/2003--Lee-B-J-Shim-J-H-Baskes-M-I--Al/}.

\section*{Code Availability}
The \texttt{chemtrain} framework, including \ctd{}, is open-source and available at \url{https://github.com/tummfm/chemtrain} with documentation avnailable at \url{https://chemtrain.readthedocs.io/en/latest/}.
The LAMMPS molecular dyamics package is publicly available at \url{https://github.com/lammps/lammps}.
Scripts for model definition, training, and benchmarking will be made publicly available at \url{https://github.com/tummfm/chemsim-lammps} upon publication of this work.

\bibliography{sn-bibliography}

\end{document}